\documentclass[english,aps,preprint,superscriptaddress]{revtex4}
\usepackage[]{fontenc}
\usepackage[latin9]{inputenc}
\usepackage{amsmath}
\usepackage{esint}
\usepackage{epsfig}
\usepackage{graphicx}
\usepackage{bm}
\usepackage{rotating}
\makeatletter
\@ifundefined{textcolor}{}
{%
 \definecolor{BLACK}{gray}{0}
 \definecolor{WHITE}{gray}{1}
 \definecolor{RED}{rgb}{1,0,0}
 \definecolor{GREEN}{rgb}{0,1,0}
 \definecolor{BLUE}{rgb}{0,0,1}
 \definecolor{CYAN}{cmyk}{1,0,0,0}
 \definecolor{MAGENTA}{cmyk}{0,1,0,0}
 \definecolor{YELLOW}{cmyk}{0,0,1,0}
 }

\makeatletter

\newcommand{\be}{\begin{equation}}\newcommand{\ee}{\end{equation}}\newcommand{\ba}{\begin{align}}\newcommand{\ea}{\end{align}}\def\bea{\begin{eqnarray}}\def\eea{\end{eqnarray}}

\usepackage{slashed}

\makeatother

\usepackage{babel}

\begin{document}

\title{Parallel Texture Structures with Cofactor Zeros in Lepton Sector}

\author{Weijian Wang}

\affiliation{Department of Physics, North China Electric Power
University, Baoding 071003, P. R. China}

\email{wjnwang96@gmail.com}

\begin{abstract}
In this paper we investigate the parallel texture structures with
cofactor zeros in the charged lepton and neutrino sectors. The
textures can not be obtained from arbitrary leptonic matrices by
making weak basis transformations, which therefore have physical
meaning. The 15 parallel textures are grouped as 4 classes where
each class has the same physical implications. It is founded that
one of them is not phenomenological viable and another is equivalent
to the texture zero structures extensively explored in previous
literature. Thus we focus on the other two classes of parallel
texture structures and study the their phenomenological
implications. The constraints on the physical variables are obtained
for each class, which are essential for the model selection and can
be measured by future experiments. The model realization is
illustrated in a radiated lepton mass model.

\vspace{1em}

\end{abstract}
\maketitle

\section{Introduction}
The discovery of neutrino oscillations has provided us with
convincing evidences for massive neutrinos and leptonic flavor
mixing with high degree of accuracy\cite{neu1,neu2,neu3}. The recent
measurement of large reactor mixing angle $\theta_{13}$ has not only
open the door for us to explore the leptonic CP violation and the
mass hierarchy in the future experiments, but also highlight the
flavor puzzle of neutrino mass and mixing pattern which appears to
be rather different from the distinct mass hierarchy and the small
mixing angles shown by quarks. Although a full theory is still
missing, several ideas have been proposed by reducing the number of
free parameters of seesaw models\cite{seesaw} and introducing the
specific structures into mass matrices to explain the observed
leptonic mixing pattern. The models include texture
zeros\cite{zero}, hybrid textures\cite{hybrid, hybrid2}, zero
trace\cite{sum}, zero determinant\cite{det}, vanishing
minors\cite{minor,minor1,minor2}, two traceless
submatrices\cite{tra}, equal elements or cofactors\cite{co}, hybrid
$M_{\nu}^{-1}$ textures\cite{hyco}. Among these models, the textures
with zero elements or zero minors are particularly interesting
because of their connection to the flavor symmetries and the stable
behavior of running renomalization group. The phenomenological
analysis of neutrino mass matrices with texture zeros or cofactor
zeros in flavor basis have been widely investigated in earlier
literature\cite{zero,minor,minor1, minor2}.

However, there is no priori requirement that the analysis must be
done in flavor basis. The more general situation should be
considered in the basis where both charged lepton mass matrix
$M_{l}$ and neutrino mass matrix $M_{\nu}$ are non-diagonal. In this
spirit, the parallel $Ans\ddot{a}tze$ has been proposed where
$M_{l}$ and $M_{\nu}$ have the same structure (We denote it
"parallel texture structure"). A popular parallel texture structure
appears as the Fritzsch-like model\cite{fritzsch} with texture zeros
in mass matrix and is firstly applied to understand the quark mixing
pattern. Subsequently the idea is generalized to the lepton
sector\cite{zzx,ylz}. A systematic search on the parallel structures
with texture zeros in lepton mass matrices are reported in
Ref.\cite{GCB}. It is shown that some sets of the texture zeros have
no physical meaning by themselves, since they can be obtained by
making suitable weak basis (WB) transformation from arbitrary mass
matrix and leaving the gauge currents invariant. The minimal no
trivial case is the four texture zeros model. Recently, a similar
investigation is done in the context of parallel hybrid textures
with one zero and two equal elements \cite{hyp}.

In this work, we study the parallel structures with two cofactor
zeros in both $M_{\nu}$ and $M_{\nu}$. As shown in Ref.
\cite{minor1}, the cofactor zeros in $M_{\nu}$ are generated by the
type-I seesaw formula $M_{\nu}=-M^{D}M_{R}^{-1}M_{D}^{T}$ with
texture zeros in $M_{D}$ and $M_{R}$. The cofactor zeros in $M_{l}$,
on the other hand, seems to be rather unusual because no flavor
symmetry directly leads to the cofactor zeros of Dirac mass matrix
$M_{l}$. However, we will show that if we adopt the recent viewpoint
proposed by Ma\cite{ma} that the radiated lepton mass originate from
the one-loop diagram, a seesaw-like formula is possible for charged
lepton masses and the cofactor zeros in $M_{l}$ can be realized.
There exists $C_{6}^2=15$ logically possible patterns. Furthermore,
we assume the mass matrices to be Hermitian and all neutrinos are
massive, which indicates det$M_{\nu}\neq 0$ and existence of
$M_{\nu}^{-1}$. Thus the mass textures $M_{\nu}$ with cofactor zeros
are equivalent to the $M_{\nu}^{-1}$ with texture zeros. As the
texture zero case \cite{GCB}, the 15 textures structures can be
grouped into 4 classes with each class having the same physical
implications. Among the 4 classes, we find that one of them is not
viable phenomenologically and another class is equal to the matrices
with texture zeros. Therefore we focus on the other two classes
having not been studied before.

The paper is organized as follows. In Sec. II, we discuss the
classification of textures and relate them to the experimental
results. In Sec.III, we diagonalize the mass matrices, confront the
numerical results with the experimental data and discuss their
predictions. In Sec. IV, the the realization of cofactor zeros in
$M_{l}$ is discussed. A summary is given in Sec. IV.

\section{Formalism}
\subsection{Weak basis equivalent classes}
We assume the neutrinos to be Majorana fermions. The most general WB
transformations leaving gauge currents invariant is given by
\begin{equation}
M_{l}\rightarrow M_{l}^{\prime}=W^{\dagger}M_{l}W_{R}\quad\quad\quad
M_{\nu}\rightarrow M_{\nu}^{\prime}=W^{T}M_{\nu}W
\end{equation}
where $W$, $W_{R}$ are $3\times 3$ unitary matrices. Therefore the
parallel texture with cofactor zeros located at different positions
can be related by permutation matrix $P$ as the WB transformation

\begin{equation}
M_{l}^{\prime}=P^{T}M_{l}P\quad\quad\quad
M_{\nu}^{\prime}=P^{T}M_{\nu}P
\end{equation}
The permutation matrix $P$ changes the positions of cofactor zeros
but preserves the parallel structures for both charged lepton and
neutrino mass textures. It is noted that $P$ belongs to the group of
6 permutations and are isomorphic to $S_{3}$. Then the four cofactor
zeros texture can be classified into 4 classes as following:

Class I:
\begin{equation}\begin{split}
\left(\begin{array}{ccc}
\bigtriangleup&\times&\bigtriangleup\\
\times&\times&\times\\
\bigtriangleup&\times&\times
\end{array}\right)\quad\quad
\left(\begin{array}{ccc}
\bigtriangleup&\bigtriangleup&\times\\
\bigtriangleup&\times&\times\\
\times&\times&\times
\end{array}\right)\quad\quad
\left(\begin{array}{ccc}
\times&\bigtriangleup&\times\\
\bigtriangleup&\bigtriangleup&\times\\
\times&\times&\times
\end{array}\right)\\
\left(\begin{array}{ccc}
\times&\times&\times\\
\times&\bigtriangleup&\bigtriangleup\\
\times&\bigtriangleup&\times
\end{array}\right)\quad\quad
\left(\begin{array}{ccc}
\times&\times&\bigtriangleup\\
\times&\times&\times\\
\bigtriangleup&\times&\bigtriangleup
\end{array}\right)\quad\quad
\left(\begin{array}{ccc}
\times&\times&\times\\
\times&\times&\bigtriangleup\\
\times&\bigtriangleup&\bigtriangleup
\end{array}\right)
\end{split}\label{matrx}\end{equation}

Class II:
\begin{equation}\begin{split}
\left(\begin{array}{ccc}
\bigtriangleup&\times&\times\\
\times&\times&\bigtriangleup\\
\times&\bigtriangleup&\times
\end{array}\right)\quad\quad
\left(\begin{array}{ccc}
\times&\times&\bigtriangleup\\
\times&\bigtriangleup&\times\\
\bigtriangleup&\times&\times
\end{array}\right)\quad\quad
\left(\begin{array}{ccc}
\times&\bigtriangleup&\times\\
\bigtriangleup&\times&\times\\
\times&\times&\bigtriangleup
\end{array}\right)
\end{split}\label{matrx}\end{equation}

Class III:
\begin{equation}\begin{split}
\left(\begin{array}{ccc}
\bigtriangleup&\times&\times\\
\times&\bigtriangleup&\times\\
\times&\times&\times
\end{array}\right)\quad\quad
\left(\begin{array}{ccc}
\bigtriangleup&\times&\times\\
\times&\times&\times\\
\times&\times&\bigtriangleup
\end{array}\right)\quad\quad
\left(\begin{array}{ccc}
\times&\times&\times\\
\times&\bigtriangleup&\times\\
\times&\times&\bigtriangleup
\end{array}\right)
\end{split}\label{matrx}\end{equation}

Class IV:
\begin{equation}\begin{split}
\left(\begin{array}{ccc}
\times&\bigtriangleup&\bigtriangleup\\
\bigtriangleup&\times&\times\\
\bigtriangleup&\times&\times
\end{array}\right)\quad\quad
\left(\begin{array}{ccc}
\times&\bigtriangleup&\times\\
\bigtriangleup&\times&\bigtriangleup\\
\times&\bigtriangleup&\times
\end{array}\right)\quad\quad
\left(\begin{array}{ccc}
\times&\times&\bigtriangleup\\
\times&\times&\bigtriangleup\\
\bigtriangleup&\bigtriangleup&\times
\end{array}\right)
\end{split}\label{matrx}\end{equation}
where "$\bigtriangleup$" at $(i,j)$ position denotes the zero
cofactor $C_{ij}=0$ while "$\times$" stands for arbitrary element.
Since $M_{l,\nu}$ with cofactor zeros is equivalent to
$M_{l,\nu}^{-1}$ with zero elements, the classification given above
is the same as the texture zero ones shown in Ref.\cite{GCB} except
for replacing "$\bigtriangleup$" with "0". Like the texture zero
cases , the class IV leads to the decoupling of a generation of
lepton from mixing and thus not experimentally viable. On the other
hand, one can easily check that the textures of class I correspond
to the texture zero ones, which has already studied in previous
literature\cite{zzx,ylz,GCB}. As an example, for the first matrix of
class I, we have
\begin{equation}\begin{split}
M_{l,\nu}=\left(\begin{array}{ccc}
\bigtriangleup&\times&\bigtriangleup\\
\times&\times&\times\\
\bigtriangleup&\times&\times
\end{array}\right)\Rightarrow
M_{l,\nu}^{-1}=\left(\begin{array}{ccc}
0&\times&0\\
\times&\times&\times\\
0&\times&\times
\end{array}\right)
\Rightarrow M_{l,\nu}=\left(\begin{array}{ccc}
\times&\times&\times\\
\times&0&0\\
\times&0&\times
\end{array}\right)
\end{split}\label{matrx1}\end{equation}
Therefore only class II and class III have no trivial physical
implications.
\subsection{Some useful notations}
As we have mentioned, we only need to investigate two mass matrices
respectively belonging to to representations of class II and class
III. In this work, we choose
\begin{equation}\begin{split}
M_{l,\nu}^{II}=\left(\begin{array}{ccc}
\bigtriangleup&\times&\times\\
\times&\times&\bigtriangleup\\
\times&\bigtriangleup&\times
\end{array}\right)\quad\quad
M_{l,\nu}^{III}=\left(\begin{array}{ccc}
\bigtriangleup&\times&\times\\
\times&\bigtriangleup&\times\\
\times&\times&\times
\end{array}\right)
\end{split}\label{matrx}\end{equation}
The charged leptonic mass texture $M_{l}$ is a complex Hermitian
matrix and the Majorana neutrino mass texture $M_{\nu}$ is a complex
symmetric matrix. They are diagonalized by unitary matrix $V_{l}$
and $V_{\nu}$
\begin{equation}
M_{l}=V_{l}M_{l}^{D}V_{l}^{\dagger}\quad\quad
M_{\nu}=V_{\nu}M_{\nu}^{D}V_{\nu}^{T}
\label{dia}\end{equation}
where $M_{l}^{D}=Diag(m_{e},m_{\mu}, m_{\tau})$,
$M_{\nu}^{D}=Diag(m_{1},m_{2},m_{3})$. The
Pontecorvo-Maki-Nakagawa-Sakata matrix\cite{PMNS} $U_{PMNS}$ is
given by
\begin{equation}
U_{PMNS}=V_{l}^{\dagger}V_{\nu} \label{upmns}\end{equation} and can
be parameterized as
\begin{equation}
U_{PMNS}=UP_{\nu}=\left(\begin{array}{ccc}
c_{12}c_{13}&c_{13}s_{12}&s_{13}e^{-i\delta}\\
-s_{12}c_{23}-c_{12}s_{13}s_{23}e^{i\delta}&c_{12}c_{23}-s_{12}s_{13}s_{23}e^{i\delta}&c_{13}s_{23}\\
s_{23}s_{12}-c_{12}c_{23}s_{13}e^{i\delta}&-c_{12}s_{23}-c_{23}s_{12}s_{13}e^{i\delta}&c_{13}c_{23}
\end{array}\right)\left(\begin{array}{ccc}
1&0&0\\
0&e^{i\alpha}&0\\
0&0&e^{i(\beta+\delta)}
\end{array}\right)
\label{3}\end{equation} where the abbreviations
$s_{ij}=\sin\theta_{ij}$ and $c_{ij}=\cos\theta_{ij}$ are used. The
$\alpha$ and $\beta$ in $P_{\nu}$ denote two Majorana CP-violating
phases and $\delta$ in U denotes the Dirac CP-violating phase. In
order to facilitate our calculation, it is better to start from
$M_{l}^{-1}$ rather than $M_{l}$. From \eqref{dia}, we get
\begin{equation}
M_{l}^{-1}=V_{l}(M_{l}^{D})^{-1}V_{l}^{\dagger}
\end{equation}
So the $V_{l}$ can not only diagonalize the $M_{l}$ but also
$M_{l}^{-1}$. Furthermore, we treat the Hermitian matrix
$M_{l}^{-1}$ to be factorisable. i.e
\begin{equation}
M_{l}^{-1}=K_{l}(M_{l}^{-1})^{r}K_{l}^{\dagger}
\end{equation}
where $K_{l}$ is the unitary phase matrix and can be parameterized
as $K_{l}=diag(1,e^{i\phi_{1}},e^{i\phi_{2}})$. The
$(M_{l}^{-1})^{r}$ becomes a real symmetric matrix which can be
diagonalized by real orthogonal matrix $O_{l}$. Then we have
\begin{equation}
V_{l}=K_{l}O_{l}
\end{equation}
and
\begin{equation}
U_{PMNS}=O_{l}^{T}K_{l}^{\dagger}V_{\nu} \label{okv}\end{equation}
From \eqref{dia}, \eqref{upmns} and \eqref{okv}, the neutrino mass
matrix $M_{\nu}$ is given by
\begin{equation}
M_{\nu}=K_{l}VP_{\nu}M_{\nu}^{D}P_{\nu}V^{T}K_{l}^{\dagger}
\label{mvd}\end{equation} where $V\equiv O_{l}U$. From \eqref{mvd}
The restriction of two cofactor zeros on $M_{\nu}$
\begin{equation}
M_{\nu(pq)}M_{\nu(rs)}-M_{\nu(tu)}M_{\nu(vw)}=0\quad\quad
M_{\nu(p^{\prime}q^{\prime})}M_{\nu(r^{\prime}s^{\prime})}-M_{\nu(t^{\prime}u^{\prime})}M_{\nu(v^{\prime}w^{\prime})}=0
\end{equation}
induces two equations
\begin{equation}
m_{1}m_{2}K_{3}e^{2i\alpha}+m_{2}m_{3}K_{1}e^{2i(\alpha+\beta+\delta)}+m_{3}m_{1}K_{2}e^{2i(\beta+\delta)}=0
\label{c1}\end{equation}
\begin{equation}
m_{1}m_{2}L_{3}e^{2i\alpha}+m_{2}m_{3}L_{1}e^{2i(\alpha+\beta+\delta)}+m_{3}m_{1}L_{2}e^{2i(\beta+\delta)}=0
\label{c2}\end{equation} where
\begin{equation}
K_{i}=(V_{pj}V_{qj}V_{rk}V_{sk}-V_{tj}V_{uj}V_{vk}V_{wk})+(j\leftrightarrow
k)\end{equation}
\begin{equation}
L_{i}=(V_{p^{\prime}j}V_{q^{\prime}j}V_{r^{\prime}k}V_{s^{\prime}k}-V_{t^{\prime}j}V_{u^{\prime}j}V_{v^{\prime}k}V_{w^{\prime}k})+(j\leftrightarrow
k)\end{equation} with $(i,j,k)$ a cyclic permutation of (1,2,3).
After solving Eq.\eqref{c1} and \eqref{c2}, we arrive at
\begin{equation}
\frac{m_{1}}{m_{2}}e^{-2i\alpha}=\frac{K_{3}L_{1}-K_{1}L_{3}}{K_{2}L_{3}-K_{3}L_{2}}
\label{r1}\end{equation}
\begin{equation}
\frac{m_{1}}{m_{3}}e^{-2i\beta}=\frac{K_{2}L_{1}-K_{1}L_{2}}{K_{3}L_{2}-K_{2}L_{3}}e^{2i\delta}
\label{r2}\end{equation}

With the help of Eq.\eqref{r1} and \eqref{r2}, we obtain the
magnitudes of mass radios
\begin{equation}
\rho=\Big|\frac{m_{1}}{m_{3}}e^{-2i\beta}\Big|
\label{t1}\end{equation}
\begin{equation}
\sigma=\Big|\frac{m_{1}}{m_{2}}e^{-2i\alpha}\Big|
\label{t2}\end{equation} as well as the two Majorana CP-violating
phases
\begin{equation}
\alpha=-\frac{1}{2}arg\Big(\frac{K_{3}L_{1}-K_{1}L_{3}}{K_{2}L_{3}-K_{3}L_{2}}\Big)
\label{21}\end{equation}
\begin{equation}
\beta=-\frac{1}{2}arg\Big(\frac{K_{2}L_{1}-K_{1}L_{2}}{K_{3}L_{3}-K_{2}L_{3}}e^{2i\delta}\Big)
\label{22}\end{equation} The results of Eq. \eqref{t1},\eqref{t2},
\eqref{21} and \eqref{22} imply that the two mass ratios ($\rho$ and
$\sigma$) and two Majorana CP-violating phases ($\alpha$ and
$\beta$) are fully determined in terms of the real orthogonal matrix
$O_{l}$ and $U$($\theta_{12}, \theta_{23}, \theta_{13}$ and
$\delta$). The neutrino mass ratios $\rho$ and $\sigma$ are related
to the ratios of two neutrino mass-squared ratios obtained from the
solar and atmosphere oscillation experiments as
\begin{equation}
R_{\nu}\equiv\frac{\delta m^{2}}{\Delta
m^{2}}=\frac{2\rho^{2}(1-\sigma^{2})}{|2\sigma^{2}-\rho^{2}-\rho^{2}\sigma^{2}|}
\label{rv}\end{equation} and to the three neutrino mass as
\begin{equation}
 m_{2}=\sqrt{\frac{\delta m^{2}}{1-\sigma^{2}}}\quad\quad
 m_{1}=\sigma m_{2}\quad\quad m_{3}=\frac{m_{1}}{\rho}
\label{abm}\end{equation} where $\delta m^{2}\equiv
m_{2}^{2}-m_{1}^{2}$ and $\Delta m^{2}\equiv \mid
m_{3}^{2}-\frac{1}{2}(m_{1}^{2}+m_{2}^{2})\mid$. In the numerical
analysis, we use the latest global-fit neutrino oscillation
experimental data, at 3$\sigma$ confidential level, which is listed
in Ref.\cite{data}
\begin{equation}\begin{split}
\sin^{2}\theta_{12}/10^{-1}=3.08^{+0.51}_{-0.49}\quad
\sin^{2}\theta_{23}/10^{-1}=4.25^{+2.16}_{-0.68}\quad
\sin^{2}\theta_{13}/10^{-2}=2.34^{+0.63}_{-0.57}\\
\delta m^{2}/10^{-5}=7.54^{+0.64}_{-0.55} eV^{2}\quad\quad
\bigtriangleup m^{2}/10^{-3}=2.44^{+0.22}_{-0.22} eV^{2}
\end{split}\end{equation}
for normal hierarchy (NH) and
\begin{equation}\begin{split}
\sin^{2}\theta_{12}/10^{-1}=3.08^{+0.51}_{-0.49}\quad
\sin^{2}\theta_{23}/10^{-1}=4.25^{+2.22}_{-0.74}\quad
\sin^{2}\theta_{13}/10^{-2}=2.34^{+0.61}_{-0.61}\\
\delta m^{2}/10^{-5}=7.54^{+0.64}_{-0.55} eV^{2}\quad\quad
\bigtriangleup m^{2}/10^{-3}=2.40^{+0.21}_{-0.23} eV^{2}
\end{split}\end{equation}
for inverted hierarchy(IH). There is no constraint on the Dirac
CP-violating phase $\delta$ at $3\sigma$ level, however, the recent
global fit tends to give $\delta\approx 1.40\pi$. In neutrino
oscillation experiments, CP violation effect is usually reflected by
the Jarlskog rephasing invariant quantity\cite{Jas} defined as
\begin{equation}
J_{CP}=s_{12}s_{23}s_{13}c_{12}c_{23}c_{13}^{2}\sin\delta
\end{equation}
The Majorana nature of neutrino can be determined if any signal of
neutrinoless double decay is observed, implying the violation of
leptonic number violation. The decay ratio is related to the
effective of neutrino $m_{ee}$, which is written as
\begin{equation}
m_{ee}=|m_{1}c_{12}^{2}c_{13}^{2}+m_{2}s_{12}^{2}c_{13}^{2}e^{2i\alpha}+m_{3}s_{13}^{2}e^{2i\beta}|
\end{equation}
 Although a $3\sigma$ result of $m_{ee}=(0.11-0.56)$ eV is
reported by the Heidelberg-Moscow Collaboration\cite{HM}, this
result is criticized in Ref \cite{NND2} and shall be checked by the
forthcoming experiment. It is believed that that the next generation
$0\nu\beta\beta$ experiments, with the sensitivity of $ m_{ee}$
being up to 0.01 eV\cite{NDD}, will open the window to not only the
absolute neutrino mass scale but also the Majorana-type CP
violation. Besides the $0\nu\beta\beta$ experiments, a more severe
constraint was set from the recent cosmology observation. Recently,
an upper bound on the sum of neutrino mass $\sum m_{i}<0.23$ eV is
reported by Plank Collaboration\cite{Planck} combined with the WMAP,
high-resolution CMB and BAO experiments.

\section{Phenomenological implications of parallel cofactor zero textures}
\subsection{Class II}
In this section, we study the phenomenological implications of class
II. The factorisable formation of inverse charged leptonic matrix
($M_{l}^{-1})^{r}$ are parameterized as
\begin{equation}\begin{split}
(M_{l}^{-1})^{r}_{II}=\left(\begin{array}{ccc}
0&a&c\\
a&b&0\\
c&0&d
\end{array}\right)
\end{split}\label{II}\end{equation}
and can be diagonalized by an orthogonal matrix $O_{l}$
\begin{equation}
O_{l}^{T}(M_{l}^{-1})^{r}_{II}O_{l}=diag(m_{e}^{-1},-
m_{\mu}^{-1},m_{\tau}^{-1}) \label{ot1}\end{equation} where the
coefficients $a, c, d$ are real and positive; The $m_{e}, m_{\mu}$
and $m_{\tau}$ denote the masse eigenvalues of charged leptons for
three generations. The minus sign in \eqref{ot1} has been introduced
to facilitate the analytical calculation and has no physical meaning
since the charged lepton is Dirac fermions. Using the invariant
Tr$(M_{l}^{-1})^{r}$, Det$ (M_{l}^{-1})^{r}$ and
Tr$(M_{l}^{-1})^{r^{2}}$ the nozero elements of $(M_{l}^{-1})^{r}$
can be expressed in terms of three mass eigenvalues $m_{e},
m_{\mu}$, $m_{\tau}$ and $d$
\begin{equation}
a=\sqrt{-\frac{(m_{e}^{-1}-m_{\mu}^{-1}-d)(m_{e}^{-1}+m_{\tau}^{-1}-d)(-m_{\mu}^{-1}+m_{\tau}^{-1}-d)}{m_{e}^{-1}-m_{\tau}^{-1}+m_{\tau}^{-1-2d}}}
\label{IIA}\end{equation}
\begin{equation}
b=m_{e}^{-1}-m_{\mu}^{-1}+m_{\tau}^{-1}-d \label{IIB}\end{equation}
\begin{equation}
c=\sqrt{\frac{(d-m_{e}^{-1})(d+m_{\mu}^{-1})(d-m_{\tau}^{-1})}{m_{e}^{-1}-m_{\tau}^{-1}+m_{\tau}^{-1}}-2d}
\label{IIC}\end{equation} where the parameter $d$ is allowed in the
range of $0<d<m_{\tau}^{-1}$ and
$m_{e}^{-1}-m_{\tau}^{-1}<d<m_{e}^{-1}$. Then the $O_{l}$ can be
easily constructed as
\begin{equation}\begin{split}
O_{l}=\left(\begin{array}{ccc}
\frac{(b-m_{e}^{-1})(d-m_{e}^{-1})}{N_{1}}&\frac{(b+m_{\mu}^{-1})(d+m_{\mu}^{-1})}{N_{2}}&\frac{(b-m_{\tau}^{-1})(d-m_{\tau}^{-1})}{N_{3}}\\
-\frac{a(d-m_{e}^{-1})}{N_{1}}&-\frac{a(d+m_{\mu}^{-1})}{N_{2}}&-\frac{a(d-m_{\tau}^{-1})}{N_{3}}\\
-\frac{c(b-m_{e}^{-1})}{N_{1}}&-\frac{c(b+m_{\mu}^{-1})}{N_{2}}&-\frac{c(b-m_{\tau}^{-1})}{N_{3}}
\end{array}\right)
\end{split}\label{II1}\end{equation}
where the $a,b$ and $c$ in\eqref{II1} is given in \eqref{IIA},
\eqref{IIB} and \eqref{IIC}; The $N_{1}$, $N_{2}$ and $N_{3}$ are
the normalized coefficients given by
\begin{equation}
N_{1}^{2}=(b-m_{e}^{-1})^{2}(d-m_{e}^{-1})^{2}+a^{2}(d-m_{e}^{-1})^{2}+c^{2}(b-m_{\tau}^{-1})^{2}
\end{equation}
\begin{equation}
N_{2}^{2}=(b+m_{\mu}^{-1})^{2}(d+m_{\mu}^{-1})^{2}+a^{2}(d+m_{\mu}^{-1})^{2}+c^{2}(b+m_{\mu}^{-1})^{2}
\end{equation}
\begin{equation}
N_{3}^{2}=(b-m_{\tau}^{-1})^{2}(d-m_{\tau}^{-1})^{2}+a^{2}(d-m_{\tau}^{-1})^{2}+c^{2}(b-m_{\tau}^{-1})^{2}
\end{equation}
Substitute the $O_{l}$ we obtained into \eqref{II1} to\eqref{t1},
\eqref{t2}, \eqref{21}, \eqref{22} and \eqref{rv}, the ratio of mass
squared difference can be expressed via eight parameters. i.e three
mixing angles $(\theta_{12}, \theta_{23}, \theta_{13})$, one Dirac
CP-violating phase $\delta$, three charged lepton mass $(m_{e},
m_{\mu}, m_{\tau})$ and a parameter $d$. Here we choose the three
charged leptonic masses at the electroweak scale($\mu\simeq M_{Z}$)
i.e\cite{zzx2}
\begin{equation}
m_{e}=0.486570154 MeV\quad\quad m_{\mu}=102.7181377MeV \quad\quad
m_{\tau}=1746.17MeV
\end{equation}
In the numerical analysis, We randomly vary the three mixing angles
$(\theta_{12}, \theta_{23}, \theta_{13})$ in their $3\sigma$ range
and parameter $d$ in its proper range. Up to now, no bound was set
on Dirac CP-violating phase $\delta$ at 3 $\sigma$ level, so we vary
it randomly in the range of $[0,2\pi]$. Using Eq. \eqref{rv}, the
mass-squared difference ratio $R_{\nu}$ is determined. Then the
input parameters is empirically acceptable when the $R_{\nu}$ falls
inside the the $3\sigma$ range of experimental data, otherwise they
are excluded. Finally, we get the value of neutrino mass and
Majorana CP-violating phase $\alpha$ and $\beta$ though
Eq.\eqref{t1}, \eqref{t2}. Since we have already obtained the
absolute neutrino mass $m_{1,2,3}$, the further constraint from
cosmology should be considered. In this work, we set the upper bound
on the sum of neutrino mass $\Sigma m_{i}$ less than 0.23 eV.

\begin{figure}
 \includegraphics[scale = 0.65]{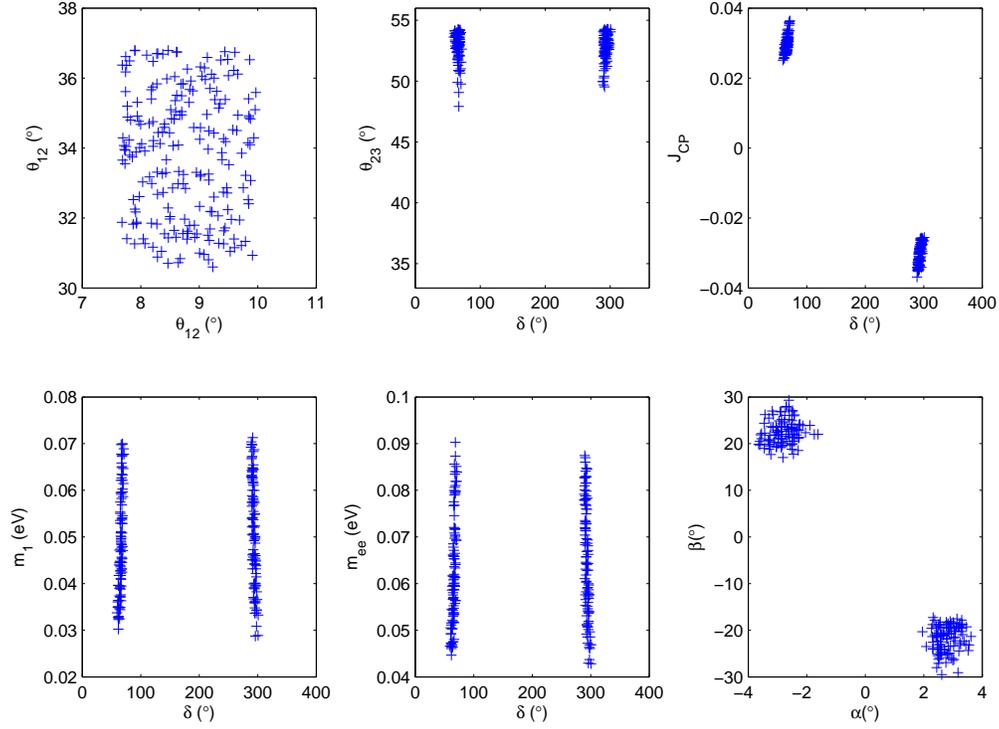}%
\caption {The correlation plots for class II(NH). } \label{IINH}
 \end{figure}

\begin{figure}
 \includegraphics[scale = 0.65]{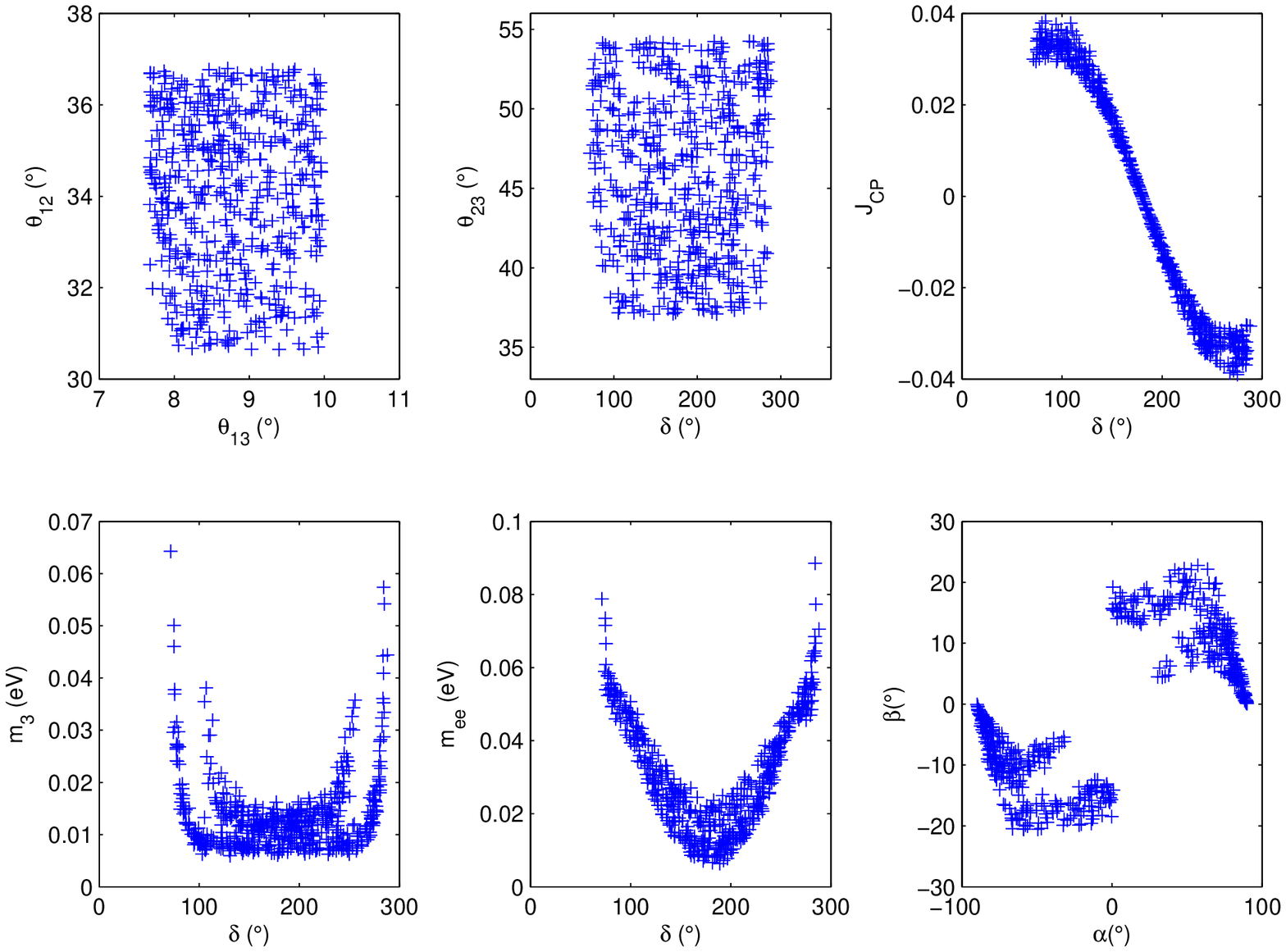}%
\caption {The correlation plots for class II(IH). } \label{IIIH}
 \end{figure}

We present the numerical results of class II in Fig.\ref{IINH} for
the NH and in Fig.\ref{IIIH} for the IH. One can see from the
figures that different mass spectra exhibit different correlations
between physical variables. For the NH case, the Dirac CP-violating
phase $\delta$ is highly restricted in the range of $60^{\circ}\sim
70^{\circ}$ and leads to the Jarlskog rephasing invariant
$|J_{CP}|>0.02$ which is promising to be detected in the future long
baseline neutrino oscillation experiments. On the other hand, there
exists a bound of $\theta_{23}>48^{\circ}$. Although accepted at
3$\sigma$ level, this result is phenomenologically ruled out at
$2\sigma$ level since recent experiments tend to give
$\theta_{23}<\pi/4$. We obtain the bound on the lightest neutrino
mass $M_{1}$, $0.025eV<m_{1}<0.075eV$ and the effective Majorana
neutrino mass $m_{ee}$ $0.04eV<m_{ee}<0.10eV$ which reaches the
accuracy of future neutrinoless double beta decay $(0\nu\beta\beta)$
experiments. The correlation between $\alpha$ and $\beta$ is also
illustrated that the small range is allowed at $3\sigma$ level. For
the IH case, the three mixing angles $\theta_{12}$, $\theta_{23}$,
and $\theta_{13}$ are fully covered the $3\sigma$ range while the
constrained Dirac CP-violated phase $\delta$ lies in the range of
$70^{\circ}\sim290^{\circ}$, leading to the $|J_{CP}|\sim
(0)-(0.04)$. Interestingly, $m_{ee}$ and the lightest neutrino mass
$m_{3}$ exhibit a strong dependence on $\delta$. Such correlations
are essential for the model selection and could be tested by
experiments. There also exists a bound of $0.005eV<m_{ee}<0.095eV$
which could be in principle tested by future $0\nu\beta\beta$
experiments. The Majorana phase $\alpha$ is covered in the whole
range of $-90^{\circ}\sim 90^{\circ}$ while $\beta$ is constrained
in the range of $-25^{\circ}\sim 25^{\circ}$.

\begin{figure}
 \includegraphics[scale = 0.65]{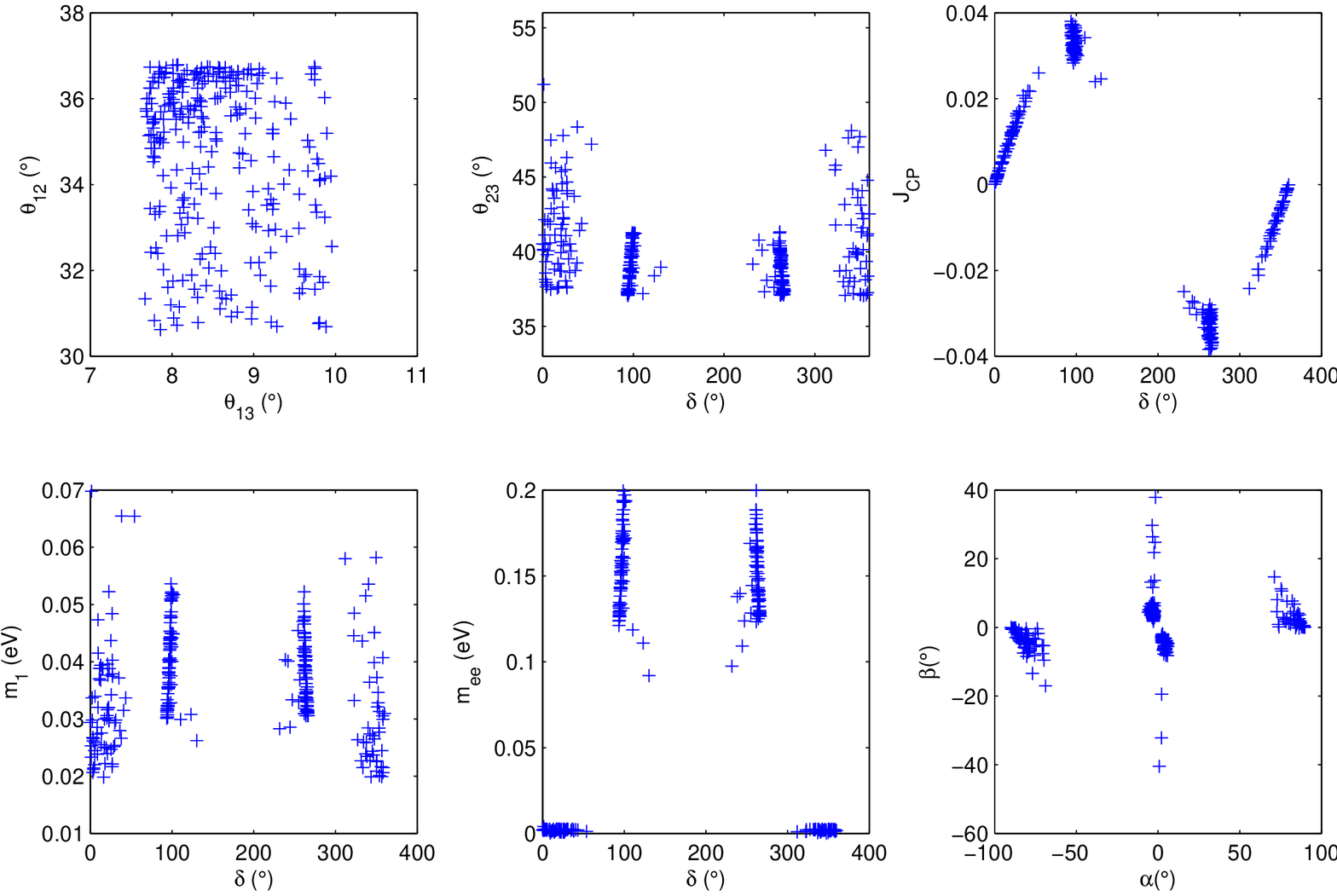}%
\caption {The correlation plots for class III(NH). } \label{IIINH}
 \end{figure}

\subsection{Class III}
Let's consider another class of textures which is phenomenologically
interesting. In the factorisable case, the real matrix
$(M_{l}^{-1})^{r}$ is parameterized as
\begin{equation}\begin{split}
(M_{l}^{-1})^{r}_{III}=\left(\begin{array}{ccc}
0&a&b\\
a&0&c\\
b&c&d
\end{array}\right)
\end{split}\label{III}\end{equation}
where $a,b,c$ and $d$ are real number. Without loss of generality,
the parameter $b,c$ are set to be positive.The matrix can be
diagonalized by the orthogonal matrix $O_{l}$
\begin{equation}
O_{l}^{T}(M_{l}^{-1})^{r}_{III}O_{l}=diag(m_{e}^{-1},-
m_{\mu}^{-1},m_{\tau}^{-1}) \label{ot}\end{equation}
Different from
class II, we choose $a$ as the free parameter since the trace of
$(M_{l}^{-1})^{r}$ has already fixed $d$ to be
\begin{equation}
d=m_{e}^{-1}-m_{\mu}^{-1}+m_{\tau}^{-1} \label{IIIab}\end{equation}
Using the invariant Det$ (M_{l}^{-1})^{r}$ and
Tr$(M_{l}^{-1})^{r^{2}}$, the parameters $b, c$ can be expressed by
three charged leptonic mass eigenvalues$(m_{e}, m_{\mu}, m_{\tau})$
and $a$
\begin{equation}
(b\pm
c)^{2}=-(-m_{e}^{-1}m_{\mu}^{-1}+m_{e}^{-1}m_{\tau}^{-1}-m_{\mu}^{-1}m_{\tau}^{-1})-a^{2}\pm
\frac{a^{2}(m_{e}^{-1}-m_{\mu}^{-1}+m_{\tau}^{-1})-m_{e}^{-1}m_{\mu}^{-1}m_{\tau}^{-1}}{a}
\label{IIIac}\end{equation} With the help of Eq. \eqref{IIIab} and
Eq. \eqref{IIIac}, one can construct the diagonalized matrix $O_{l}$
to be
\begin{equation}\begin{split}
(M_{l}^{-1})^{r}_{III}=\left(\begin{array}{ccc}
\frac{O(11)}{N_{1}}&\frac{O(12)}{N_{2}}&\frac{O(13)}{N_{3}}\\
\frac{O(21)}{N_{1}}&\frac{O(22)}{N_{2}}&\frac{O(23)}{N_{3}}\\
\frac{O(31)}{N_{1}}&\frac{O(32)}{N_{2}}&\frac{O(33)}{N_{3}}
\end{array}\right)
\end{split}\label{III}\end{equation}
where
\begin{equation}\begin{split}
O(11)=&a
m_{e}(bm_{e}+ca^{-1})+bm_{e}(m_{e}^{-1}a^{-1}-m_{e}a)\\
O(12)=&-a
m_{\mu}(-bm_{\mu}+ca^{-1})-bm_{\mu}(-m_{\mu}^{-1}a^{-1}+m_{\mu}a)\\
O(13)=&a
m_{\tau}(bm_{\tau}+ca^{-1})+bm_{\tau}(m_{\tau}^{-1}a^{-1}-m_{\tau}a)\\
&O(21)=bm_{e}+ca^{-1}\\
&O(22)=-bm_{\mu}+ca^{-1}\\
&O(23)=bm_{\tau}+ca^{-1}\\
&O(31)=m_{e}^{-1}a^{-1}-m_{e}a\\
&O(32)=-m_{\mu}^{-1}a^{-1}+m_{\mu}a\\
&O(33)=m_{\tau}^{-1}a^{-1}-m_{\tau}a\\
\end{split}\end{equation}
and the normalized coefficients is given by
\begin{equation}\begin{split}
N_{1}^{2}=O(11)^{2}+O(21)^{2}+O(31)^{2}\\
N_{2}^{2}=O(12)^{2}+O(22)^{2}+O(32)^{2}\\
N_{3}^{2}=O(13)^{2}+O(23)^{2}+O(33)^{2}
\end{split}\end{equation}
From the condition that $b,c$ are real and positive, we have the
free parameter $a$ allowed in the range of
\begin{equation}
-\Big(\frac{m_{e}^{-1}m_{\mu}^{-1}m_{\tau}}{m_{e}^{-1}-m_{\mu}^{-1}+m_{\tau}^{-1}}\Big)^{\frac{1}{2}}<a<0
\end{equation}
or
\begin{equation}
\Big(\frac{m_{e}^{-1}m_{\mu}^{-1}m_{\tau}}{m_{e}^{-1}-m_{\mu}^{-1}+m_{\tau}^{-1}}\Big)^{\frac{1}{2}}<a<(m_{e}^{-1}m_{\mu}^{-1}+m_{\mu}m_{\tau}-m_{e}^{-1}m_{\tau}^{-1})^{\frac{1}{2}}
\end{equation}

Now we repeat the previous analysis. The class III with inverted
hierarchy are now found to be unacceptable by current experimental
data. We present the allowed region for class III with normal mass
hierarchy in Fig.\ref{IIINH}. It is observed that no bound is set on
$\theta_{12}$ and $\theta_{13}$. However the Dirac CP-violating
phase $\delta$ is restricted in two regions. We denote them
respectively as R1:
$0^{\circ}<\delta<60^{\circ}(300^{\circ}<\delta<360^{\circ})$ and
R2: $90^{\circ}<\delta<150^{\circ}(210^{\circ}<\delta<270^{\circ})$.
Each shows different predictions. In R1, $\theta_{23}$ varies in its
$3\sigma$ range. We obtain a highly suppressed $m_{ee}\simeq 0$eV
which is beyond the accuracy of future $0\nu\beta\beta$ experiments
and implies the underlying cancelation of three neutrino masses in
$m_{ee}$. There also exists the lower bound on the lightest neutrino
mass $m_{1}>0.02eV$. On the other hand, in R2 we find
$\theta_{23}<45^{\circ}$ which is supported by $2\sigma$
experimental constraint. We also obtain $|J_{CP}|>0.03$. Moveover,
the bound on $m_{ee}$ is founded in the range of $0.075$eV $\sim
0.2$eV and can be potentially detected by future experiment.

\begin{figure}
 \includegraphics[scale = 0.6]{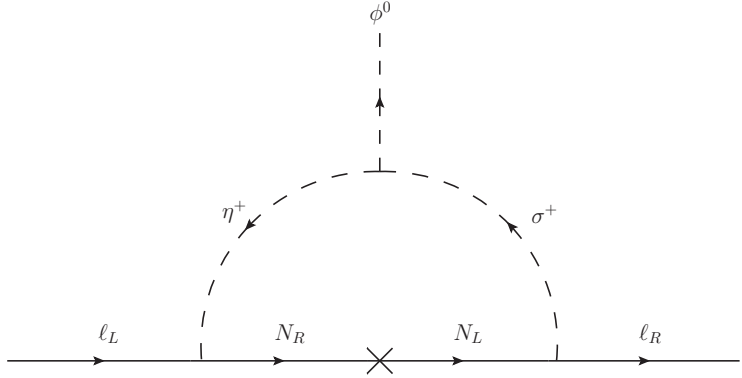}%
\caption {The one-loop diagram for generating radiated charged
lepton masses} \label{0006}
 \end{figure}

\section{Cofactor zeros in charged lepton matrices}
One reminds the type-I seesaw mechanism as
$M_{\nu}=-M_{D}M_{R}M_{D}^{T}$. Then the cofactor zeros of $M_{\nu}$
are attributed to the texture zeros in $M_{D}$ and $M_{R}$.
Generally, this can be easily realized by Abelian $Z_{n}$ flavor
symmetry\cite{zn,minor1}. Can the cofactor zeros in $M_{l}$ arise
using the same way? At the tree level, it is obviously impossible.
At the loop level, the answer is yes! Here we adopt the model
proposed by Ma\cite{ma}, consisting of the SM extended by adding
three Dirac singlet neutral fermion $N_{k} (k=1,2,3)$, a doublet
scalar $(\eta^{+}, \eta^{0})$ and a charged singlet $\sigma^{+}$. In
Ma's model, the particles transforms under the proper $U(1)_{D}$
gauge symmetry and $A_{4}$ flavor symmetry. Here, we choose the
$Z_{2}^{(A)}$ instead of $A_{4}$ flavor symmetry under which
$N_{k}$, $(\eta^{+}, \eta^{0})$ and $\sigma^{+}$ are odd. To forbid
the tree level Dirac lepton mass, another $Z_{2}^{(B)}$ symmetry is
imposed such that $l_{R}$ and $\sigma^{+}$ are odd while others are
even. Actually, the flavor symmetry we propose is the same as the
one in Ref\cite{ma1} where the Dirac neutrino mass is generated at
one-loop level. The allowed Yukawa interactions are
$y_{ij}\overline{N}_{iR}(l_{jL}\eta^{+}-\nu_{jL}\eta^{0})$ and
$h_{ij}\overline{l}_{iR}N_{jL}\sigma^{-}$. The $Z_{2}^{(B)}$ is
allowed to be softly broken by the trilinear term $\mu
(\eta^{+}\phi^{0}-\eta^{0}\phi^{+})\sigma^{-}$ with the SM vacuum
expectation $v=\langle\phi^{0}\rangle$. The one-loop charged lepton
mass is thus generated as shown in Fig. \ref{0006}, the result being
\begin{equation}
(M_{l})_{ij}=\frac{\sin 2\theta}{32\pi^{2}}\sum\limits_{k}
y_{ik}M_{k}\Big[ \frac{m_{1}^{2}}{m_{1}^{2}-M_{k}^{2}}\ln\Big(
\frac{m_{1}^{2}}{M_{k}^{2}}\Big)-\frac{m_{2}^{2}}{m_{2}^{2}-M_{k}^{2}}\ln\Big(
\frac{m_{2}^{2}}{M_{k}^{2}}\Big)\Big]h_{kj}^{\dag}
\label{mass}\end{equation} The $m_{1,2}$ and $\theta$ denote the
eigenvalues and the mixing angle of mass squared texture
\begin{equation}\begin{split}
\left(\begin{array}{cc}
m_{\sigma}^{2}&\mu v\\
\mu v&m_{\eta}^{2}
\end{array}\right)
\end{split}\end{equation}
with
\begin{equation}
m_{1,2}^{2}=\frac{1}{2}\big[m_{\eta}^{2}+m_{\sigma}^{2}\mp
\sqrt{(m_{\eta}^{2}-m_{\sigma}^{2})^{2}+4\mu^{2}v^{2}}\big ]
\end{equation}
and
\begin{equation}
\sin2\theta=\frac{2\mu^{2}v^{2}}{\sqrt{(m_{\eta}^{2}-m_{\sigma}^{2})^{2}+4\mu^{2}v^{2}}}
\end{equation}

For $M_{k}\gg m_{1,2}$, \eqref{mass} is simplified as
\begin{equation}
(M_{l})_{ij}\simeq \frac{\sin 2\theta}{32\pi^{2}} m_{1}^{2}
\sum\limits_{k} F\Big(\frac{m_{1}^{2}}{m_{2}^{2}},
\frac{M_{k}^{2}}{m_{1}^{2}}\Big) y_{ik}\frac{1}{M_{k}}h^{\dag}_{kj}
\end{equation}
with
\begin{equation}
F(x,y)\equiv x \ln \Big(\frac{y}{x}\Big)+\ln y
\end{equation}
Following the same strategy of Ref.\cite{gu}, the
$F\Big(\frac{m_{1}^{2}}{m_{2}^{2}},
\frac{M_{k}^{2}}{m_{1}^{2}}\Big)$ is treated as a constant at
leading order if three $M_{k}$ are assumed to be nearly degenerated.
Then we get
\begin{equation}
M_{l}\sim m_{1} y (M_{N})_{diag}^{-1} h^{\dag}
\label{mass1}\end{equation} On the other hand, if we assume
$m_{\eta}\simeq m_{\sigma}\simeq M_{k}$ and note $\mu v\ll
M_{k}^{2}$ , then
\begin{equation}
(M_{l})_{ij}\simeq \frac{\mu v}{16\pi^{2}}\sum
\limits_{k}y_{ik}\frac{1}{M_{k}}h^{\dag}_{kj}\sim \mu v y
(M_{N})_{diag}^{-1} h^{\dag} \label{mass2}\end{equation} The
expression also appears in \cite{ma3} where the Majorana neutrino
mass is generated at one-loop level. From \eqref{mass1} and
\eqref{mass2}, the charged leptons acquire the radiated masses via
the seesaw-like mechanism and masses of heavy Dirac neutral
particles $N_{k}$ play the role of seesaw scale.

Consider now the weak basis where the mass matrix of $M_{N}$ is not
diagonal. It is obvious that, working in the context of the
seesaw-like mechanism with a diagonal Dirac matrices $y$ and $h$,
the vanishing cofactors in the charged lepton mass matrix are
equivalent to texture zeros in the heavy Dirac fermion mass matrix
$M_{N}$. As having done in neutrino sector, the texture zeros in
$y$, $h$, and $M_{N}$ are easily achieved by introducing extra
$Z_{n}$ flavor symmetries. Form eq. \eqref{mass2}, it is clear that
the seesaw-like scale $M_{N}$ is reduced to TeV by the smallness of
factor $\mu v$/$16\pi^{2}$ originated from softly broken
$Z^{(B)}_{2}$.

\section{Conclusion and discussion}
In this work, we have studied the parallel structures with cofactor
zeros in lepton mass matrices. These matrices can not obtained from
arbitrary Hermitian texture by making WB transformations. Using the
permutation transformation, the 15 possible textures are grouped
into 4 classes where the matrices in each class lead to the same
physical implications. Among the 4 classes, one of them is not
compatible with experimental results and another is equivalent to
the texture zero structures explored extensively in previous
literature. We focus on the other two classes (class II and class
III). Using the new results from the neutrino oscillation and
cosmology experiments, a systematic and phenomenological analysis
are proposed for each class and mass hierarchy. We have demonstrated
that some predictions for the atmosphere mixing angle $\theta_{23}$,
the Dirac CP-violating phase $\delta$ and the Majorana effective
neutrino mass $m_{ee}$ are rather interesting and deserve to be
explored in the future experiments. We also demonstrate how the
cofactor zeros arise in a seesaw-like model where charged lepton
mass are generated at one-loop level. We expect that a cooperation
between phenomenological study and the flavor symmetry study will
finally help us real the structure of leptonic texture.
\begin{acknowledgments}
The author would like to thank Z. Z. Xing for the useful discussion
during this work and J. Zhang for his kind help in plotting Fig.
\ref{0006}.
\end{acknowledgments}

\end{document}